\documentclass[journal]{IEEEtran}
\ifCLASSINFOpdf
\else
\fi
\usepackage{graphicx}
\usepackage{amsmath,amsfonts, amssymb}
\usepackage{booktabs}
\usepackage{array}
\usepackage{hyperref}
\usepackage{url}
\usepackage{algorithm}
\usepackage{algpseudocode}
\usepackage{mathtools}
\usepackage{graphicx}
\usepackage{subfig}

\begin{document}
%
\title{Localization and Pursuit of a Mobile Target using Distance-only Measurements}
%
%
%

\author{Nabarupa Das,
        Suvadip Batabyal
\thanks{N. Das and S. Batabyal are with the Department of Computer Science \& Engineering
at National Institute of Technology, Durgapur,
India. e-mail: nd.25cs1110@nitdgp.ac.in, ~sbatabyal.cse@nitdgp.ac.in}}

%
%

\markboth{Journal of \LaTeX\ Class Files,~Vol.~14, No.~8, August~2015}%
{Shell \MakeLowercase{\textit{et al.}}: Bare Demo of IEEEtran.cls for IEEE Journals}
%



\maketitle

\begin{abstract}
This paper investigates localization and pursuit of a linearly moving target in a two-dimensional plane using distance measurements only. The distance from the target is estimated from pathloss measurements and is assumed to be noise-free. The proposed system consists of a receiver that is stationed at the origin and a mobile agent. The system does not rely on GPS, angle sensing, or multiple anchor nodes to localize the target. The proposed method first identifies the target quadrant and estimates its position. Consecutive target position estimates are then used to determine the motion vector and update the agent trajectory during pursuit. We show that a maximum of thirteen steps are required to localize the target and determine its motion vector. Simulation results show that the agent can successfully make a transition from search to pursuit phase, and also maintain a bounded tracking error.
\end{abstract}


\begin{IEEEkeywords}
Localization, pursuit, mobile target, pathloss, distance measurements.

\end{IEEEkeywords}

%
\IEEEpeerreviewmaketitle

\section{Introduction}
%
%
%
%

\IEEEPARstart{L}{OCALIZATION} and target pursuit are important problems in wireless sensing, autonomous and remote monitoring, and surveillance applications. In many practical scenarios, a mobile agent must follow a mobile target without knowing its exact position or location or other geometric information, visual sensing, or without the availability of multiple anchors. For example, in military surveillance, the aim of the pursuing agent is to be able to either intercept the intruder and destroy it, or be sufficiently near it to dismantle it. There exist several range-based methods to localize and pursue a target of interest such as using received signal strength indicator (RSSI), angle-of-arrival (AoA), time-of-arrival (ToA) and time-difference of arrival (TDoA) \cite{sur1}. Such methods are useful because they can be obtained from ordinary wireless signals without involving additional ranging hardware (such as radars), in GPS-denied environments, or where visual sensors are not available. 

Most conventional RSSI-based localization methods use multiple static reference nodes or anchor nodes \cite{rssi}. On the other hand, AoA method either requires directive antennas or antenna arrays to measure the direction of the arriving signal \cite{aoa1}\cite{aoa2}. Such methods are not directly suitable for a minimal target-following setup with only one fixed receiver and one mobile agent. The proposed method therefore considers distance-only information to localize and pursue a moving target. This not only overcomes the drawbacks of range-based methods mentioned above but is also simple, fast, and incurs minimal overhead. Previous studies on range-only tracking also show that observer motion and target location affect observability and tracking performance \cite{dist2}. Moreover, many UAV-based target tracking methods assume multiple beacons, cameras, range-rate information, or richer sensing conditions \cite{dist1}. Bopardikar \emph{et al.} \cite{purs1} proposed a discrete-time, pursuit-evasion game with an upper limit on the step-size of the evader. A gossip-based distributed kalman filtering approach was proposed for tracking a moving target using heterogeneous agents with limited sensing and communication ranges \cite{ma2016gossip}. However, it requires a connected multi-agent network and assumes that gossip consensus is achieved within each measurement interval. Chaudhary \emph{et al.} \cite{purs2} studied the capture of stationary and moving targets using only range and range-rate measurements, without requiring bearing information or target-position estimation. They derived sufficient conditions for the robot control law and validated the proposed approach through simulations with measurement noise and experiments using mobile robots. However, the method requires both range and range-rate information, is sensitive to range-rate noise, and only guarantees capture within a predefined radius without estimating the exact target position or motion vector. A method closely related to the proposed method is the RSSI-based law-of-cosines tracking method \cite{hunt}. In this work, successive RSSI measurements are converted into range estimates, and two consecutive ranges, together with the known movement of the mobile agent, are used to form a triangle. The law of cosines is then applied to determine a steering angle that guides the agent toward the moving target through repeated updates.

We consider the problem of a distance-only localization and pursuit problem of a mobile target moving linearly with constant velocity. For this, we use a fixed receiver and a mobile agent that can cooperatively make decisions to localize the target in a two-dimensional Euclidean plane. The proposed method uses distance only information that is derived from pathloss measurements. We employ a two phase search method first to determine the instantaneous quadrant of the target and then determining its motion vector. Subsequently, we calculate its linear velocity and using the direction of motion from the previous step we pursue the target. Unlike the proposed method, \cite{hunt} determines only the steering direction for the Hunter to move it closer to the Prey rather than localize it. The authors also assume that the Hunter will always turn counter-clockwise with angle determined using the law of cosines. Since the prey is moving with random walk mobility model, the Hunter will turn half of the time toward the Prey and the rest of the times toward the other half-plane. As a result, it does not guarantee that the Hunter intercepts target position within a fixed number of steps and depends on the initial distance between the hunter and prey and velocity of the target. In contrast, the proposed method determines the exact target coordinates and motion vector before pursuit. Hence, there exists a fixed tracking error and an upper bound on the convergence (section 4).




\section{System Model}
\noindent The system consists of a mobile target ($T$) moving in a straight line with a constant velocity $\nu$ meter/second that is to be localized and pursued by a mobile agent ($A$). The target has an isotropic transmitter. The signal from the target is received by a fixed receiver ($R$) stationed at the origin $(0,0)$, as well as the mobile agent that can move at any arbitrary speed. The system is modeled as a discrete-time system with $R$ and $A$ estimating the distance from $T$ at fixed time-intervals. It is also assumed that $R$ and $A$ can exchange simple messages such as distance estimates in real-time.

\begin{figure}
    \centering
    \includegraphics[width=0.6\linewidth]{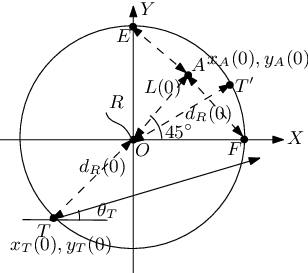}
    \caption{Geometric representation of the system model}
    \label{fig:placeholder}
\end{figure}

Let at time-instant $k$, the transmitter and agent positions be denoted by $p_T(k)=(x_T(k), y_T(k))$ and $p_A(k)=(x_A(k), y_A(k))$, respectively.
Since the transmitter moves with a constant velocity  $\nu_T$, the instantaneous position of $T$ is given by
\begin{align}
x_T(k+1) &= x_T(k) + \nu \cos\theta_T\\
y_T(k+1) &= y_T(k) + \nu \sin\theta_T
\end{align}
where, $\theta_T$ is the angle measured with respect to the positive $x$-axis. Without the loss of generality, we also assume that distance estimates from path loss measurements are noiseless and left for future work.
At each time instant $R$ and $A$ calculates the pathloss for the signal received from $T$. We assume that $T$ is always in line-of-sight (LoS) with $R$ and $A$; hence, the free-space pathloss can be calculated as 
\begin{equation}
P_r(d)=P_r(d_0)-10\eta\log_{10}\left(\frac{d}{d_0}\right)
\end{equation}
where, $P_r(d_0)$ denotes the reference signal strength at distance $d_0$, and $\eta$ is the pathloss exponent.
Therefore, the pathloss-derived ranges are treated as exact distance estimates. We denote the transmitter-to-receiver and transmitter-to-agent distances by $d_R(k)$ and $d_A(k)$, respectively. Since $R$ is at the origin, the receiver-to-agent distance at time instant $k$ is given by
\begin{equation}
L(k)=\sqrt{x_A^2(k)+y_A^2(k)}
\end{equation}
The tracking error is defined as
\[
e(k)=\left\|p_T(k)-p_A(k)\right\|
\]
The objective is to guide the aerial agent to determine the motion vector of the target and then pursue it using the distance information derived from the pathloss measurements. We also assume that the upper bound of $\nu_A$ and $\nu_T$ are same, such that $A$ is always able to catch $T$.


\begin{figure}[ht]
    \centering
    \includegraphics[width=0.8\linewidth]{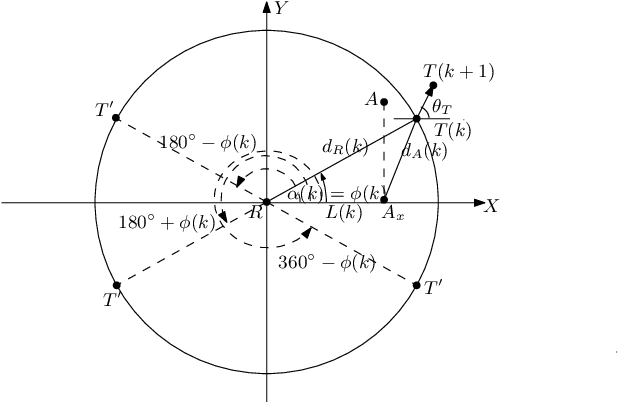}
    \caption{Geometric representation of target trajectory}
    \label{fig:motion_vector}
\end{figure}
\section{Proposed Method}
\noindent The proposed framework has three phases: the search phase, motion vector computation and the pursuit phase. The search phase identifies the target $T$ quadrant, motion vector computation phase determines the motion vector of the target $T$ that is, the velocity and the angle of motion. The objective of the pursuit phase is to estimate the instantaneous position of $T$, and update the position of $A$ to minimize $e(k)$.

\subsection{Search Phase}
\noindent The objective of the search phase is to ultimately align $A$ in the same quadrant as $T$ and then determine $T's$ motion vector. Let us assume that $T$ starts its motion from location $p_T(0)=\{x_T(0),~y_T(0)\}$. As mentioned earlier, $T$ beacons at regular intervals that is received by $R$ and $A$. From this, $R$ and $A$ estimates the distance $d_R(0)$ and $d_A(0)$, respectively. We initially place $A$ at location $p_A(0)$ given by
\begin{align}
x_A(0) &= d_R(0)\cos45^\circ-c\\
y_A(0) &= d_R(0)\sin45^\circ-c
\end{align}
for $c>0$, such that $A$ is initially placed on the first quadrant inside the circle with radius $d_R(0)$. Since at each time-instant $R$ measures only distance $d_R(.)$, t is imperative that $T$ may lie anywhere on the circumference with radius $ d_R(.)$. Similarly, $A$ also measures $d_A(0)$ with respect to $T$. In order to determine if $A$ is in the same quadrant as $T$, it measures the distance $AE$ and $AF$, respectively, where $E$ and $F$ are the points where the circle intersects the X and Y axes, respectively. Therefore, $AE=\sqrt{(y_A(0)-d_R(0))^2+x_A(0)^2}$ and $AF=\sqrt{y_A(0)^2+(x_A(0)-d_R(0))^2}$. It can now be inferred that if $AE<d_A(0)$ or $AF<d_A(0)$, then $A$ and $T$ are not in the same quadrant; else, $A$ and $T$ will be in the same quadrant. If $A$ and $T$ are found to be in the same quadrant, the next step is to determine the motion vector of $T$; otherwise, $A$ is shifted to the next (Second quadrant) and the same method is repeated until $A$ and $T$ are in the same quadrant. Once $A$ and $T$ are in the same quadrant, the next step is to determine the motion vector of $T$. For this we store the current quadrant to be used in the next step.

\subsection{Motion vector computation}
\noindent The objective of the motion vector computation is to determine the angle of motion and velocity of the target. For simplicity, we assume that $T$ does not change its quadrant until the motion vector is computed. This problem will be considered in the future version of the problem. The agent is now aligned with the x-axis with new position denoted as $A_x$ as shown in figure \ref{fig:motion_vector}. In principle, the agent may be aligned with either x-axis or y-axis; however, in this work the $x$-axis alignment is considered to simplify the geometrical construction. After alignment, $R$, $A_x$, and $T$ form the triangle $\triangle RA_xT$ as shown in Fig.~\ref{fig:motion_vector}.

Here, $R$ is the fixed receiver located at the origin, $A_x$ is the agent positioned on the x-axis, and $T$ is the target with position $p_T(k)$. As usual, $R$ and $A_x$ estimates the distance of $T$ from the pathloss measurements. Additionally, $A_x$ can also determine its distance from $R$ since it knows its own location. Therefore, the three sides of the triangle are known: the transmitter-to-receiver distance $d_R(k)$, transmitter-to-agent distance $d_A(k)$ and receiver-to-agent distance $L(k)$. Since the lengths of the three side are known, the interior angle at the receiver, $\phi(k)=\angle TRA_x$, can be computed using law of cosines. The angle $\phi(k)$ is computed as

\begin{equation}
\phi(k)=\cos^{-1}
\left(
\frac{d_R^2(k)+L^2(k)-d_A^2(k)}
{2d_R(k)L(k)}
\right)
\end{equation}

To determine the actual direction of $T$ in the two-dimensional plane, the stored quadrant during the search phase is used. Therefore, $\phi(k)$ is mapped according to the stored quadrant, and the direction angle $\alpha(k)$ of $T$ with respect to $R$ is obtained as
\begin{equation}
\alpha(k) =
\begin{cases}
\phi(k) & \text{if } T \text{ lies in Quadrant I}\\
180^\circ-\phi(k) &  \text{if } T \text{ lies in Quadrant II}\\
180^\circ+\phi(k) & \text{if } T \text{ lies in Quadrant III}\\
360^\circ-\phi(k) & \text{if } T \text{ lies in Quadrant IV}
\end{cases}
\end{equation}
Once the angle is obtained, the transmitter coordinates are reconstructed:
$x_T=d_R\cos\alpha(k)$, $y_T=d_R\sin\alpha(k)$.
To compute the motion vector of the target, two consecutive target positions are required. Let the target positions determined at time $k$ and $k+1$ be $(x_T(k),~y_T(k))$ and $(x_T(k+1),~y_T(k+1))$, respectively. Using the two consecutive positions, the distance moved by $T$ per time unit (velocity) is computed as
\begin{equation}
d_T =
\sqrt{
(x_T(k+1)-x_T(k))^2+
(y_T(k+1)-y_T(k))^2
}
\end{equation}

and the direction of motion is obtained as
\begin{align}
\theta_T&=\operatorname{atan2}
\big(y_T(k+1)-y_T(k),
x_T(k+1)-x_T(k)\big)
\end{align}
The function $\operatorname{atan2}$ is used because it considers the signs of both coordinate differences and therefore identifies the correct direction of motion in all four quadrants.
\subsection{Pursuit Phase}
\noindent The pursuit phase starts after the motion vector of the target is computed. In this phase, $A$ uses the velocity and direction of $T$ to update its own position. Let the current position of the agent be $p_A(k)=(x_A(k),y_A(k))$. Since the target moves by distance $d_T$ along the direction $\theta_T$, the next position of the agent is computed as
\begin{align}
x_A(k+1)&=x_A(k)+d_T(k)\cos\theta_T\\
y_A(k+1)&=y_A(k)+d_T(k)\sin\theta_T
\end{align}
Thus, once the motion vector of $T$ is obtained, the agent updates its position using the same displacement information and follows the target trajectory.

\section{Analysis}

\subsection{Correctness Analysis}
\noindent The proposed method has the following assumptions:

\textbf{Assumption 1:} The range information derived from the pathloss measurements is noise-free.

\textbf{Assumption 2:} The target $T$ moves with constant velocity during the observation interval.

\textbf{Assumption 3:} After the search phase is completed, the target $T$ does not immediately cross into another quadrant before its motion vector is determined.


\emph{Lemma 1:} After the pursuit phase starts, if the agent $A$ is updated using the determined motion vector of $T$, then the tracking error remains constant.

\emph{Proof:}
Let $k_L$ denote the time instant at which the motion vector of $T$ has been determined and the pursuit phase starts. Let the displacement of $T$ in one time step be denoted by $\Delta p_T$. Since $T$ moves with constant velocity, for all $k\geq k_L$,
\begin{equation}
p_T(k+1)=p_T(k)+\Delta p_T 
\end{equation}

In the pursuit phase, the agent $A$ is updated using the same determined displacement. Therefore,
\begin{equation}
p_A(k+1)=p_A(k)+\Delta p_T 
\end{equation}

Subtracting (14) from (13)
\begin{equation}
p_T(k+1)-p_A(k+1)=p_T(k)-p_A(k)
\end{equation}

Taking the norm on both sides, we obtain
\begin{equation}
\|p_T(k+1)-p_A(k+1)\|
=
\|p_T(k)-p_A(k)\|
\end{equation}

We denote the tracking error as
\begin{equation}
e(k)=\|p_T(k)-p_A(k)\|
\end{equation}

From (16) and (17) we get
\begin{equation}
e(k+1)=e(k), \qquad k\geq k_L 
\end{equation}



Therefore, the tracking error remains constant after the pursuit phase starts.
\hfill $\blacksquare$

\begin{figure*}[t]
    \centering
    \subfloat[Case 1]{
        \includegraphics[width=0.48\columnwidth]{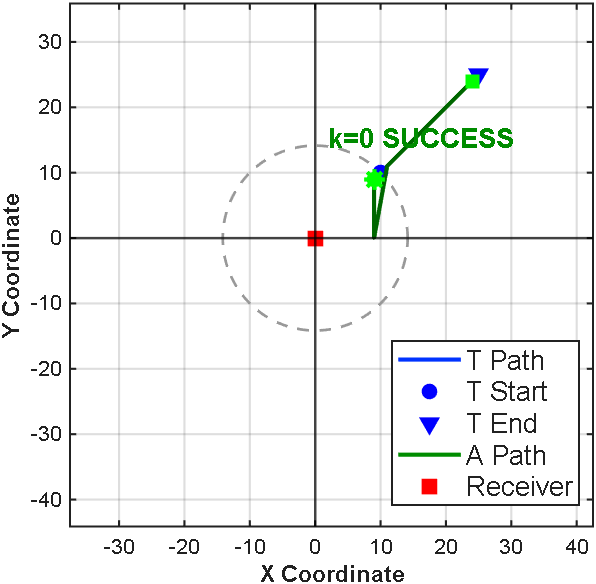}
        \label{fig:case1}
    }
    \subfloat[Case 2]{
        \includegraphics[width=0.48\columnwidth]{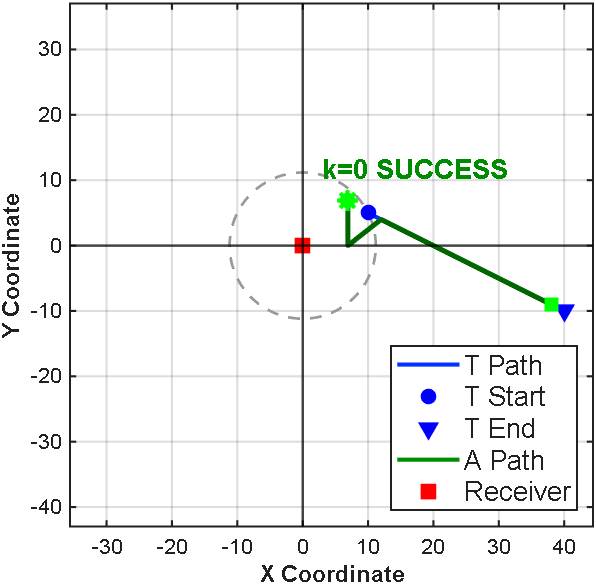}
        \label{fig:case2}
    }
    \subfloat[Case 3]{
        \includegraphics[width=0.48\columnwidth]{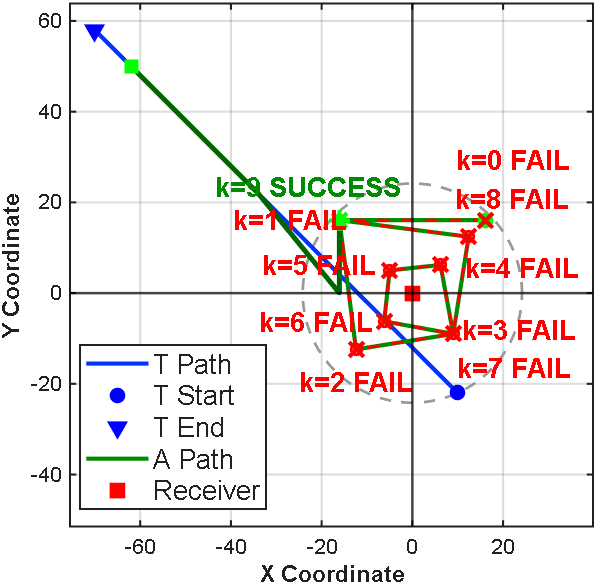}
        \label{fig:case3}
    }
    \subfloat[Case 4]{
        \includegraphics[width=0.48\columnwidth]{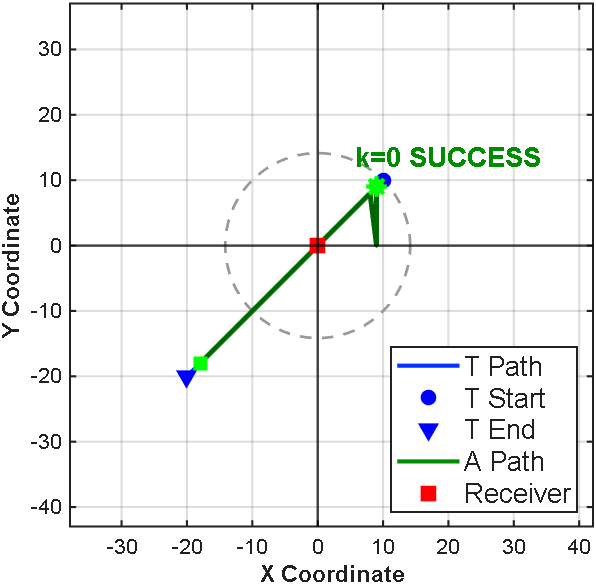}
        \label{fig:case4}
    }

    \subfloat[Case 5]{
        \includegraphics[width=0.48\columnwidth]{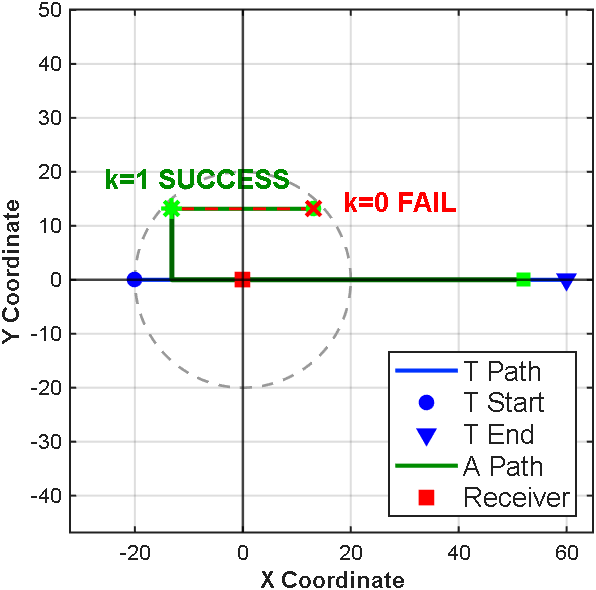}
        \label{fig:case5}
    }
    \subfloat[Case 6]{
        \includegraphics[width=0.48\columnwidth]{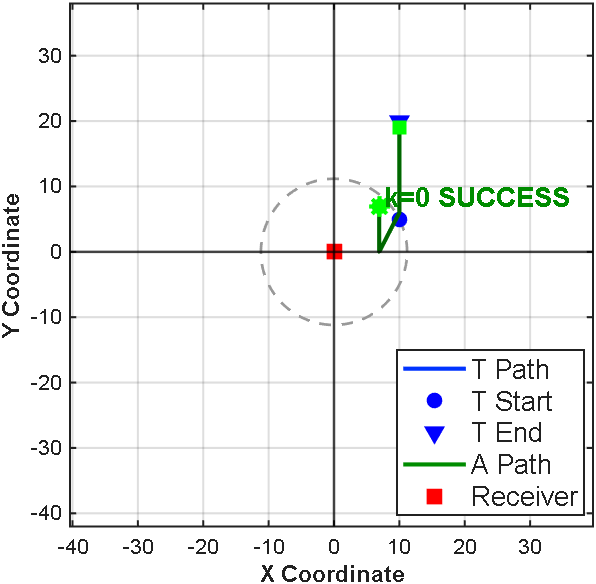}
        \label{fig:case6}
    }
    \subfloat[Case 7]{
        \includegraphics[width=0.48\columnwidth]{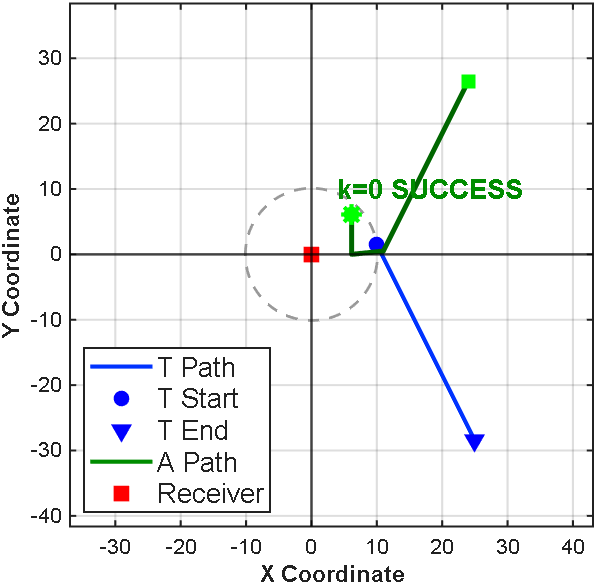}
        \label{fig:case7}
    }
    \caption{Simulation results for two target-motion cases.}
\end{figure*}
\subsection{Convergence analysis } 
\noindent We determine the upper-bound on the number of steps required to determine the motion vector of the target and the point from which the tracking error becomes constant. Let $N_s$ denote the number of steps required in the search phase. In the proposed scheme, the best-case value of $N_s$ is 1; that is, when $A$ detects $T$ in the same quadrant in the very first step. This occurs when $A$ and $T$ start in the same quadrant. Hence,$N_s^{\min}=1$. The worst-case can occur under the following \emph{explicit} scenario: (i) for anti-clockwise search: when $A$ starts at $l^{th}$ quadrant, $T$ starts at quadrant $((l+2)\mod 4)+1$, or for clockwise search: when $A$ starts at $l^{th}$ quadrant, $T$ starts at quadrant $((l\mod 4 )+1)$, (ii) $T$ traverses through 3 different quadrants, and (iii) $T$ changes its quadrant as soon as $A$ (re)appears in the same quadrant during the search phase. The worst-case scenario is explained with an example. Suppose $A$ starts from quadrant 1 ($l=1$), denoted as $Q1$, and performs an anti-clockwise search, and $T$ starts at quadrant $((1+2)\mod 4)+1)=4$, denoted as $Q4$. However, before the agent reaches $Q4$, the $T$ moves from $Q4$ to $Q3$. Therefore, $A$ misses $T$ in the fourth search. Similarly, when the agent reaches $Q3$ in the seventh step, the target moves to $Q2$. Finally, when the agent reaches $Q2$ in the tenth step, both the agent and the target are now in the same quadrant. Thus, the maximum number of steps in the search phase is$N_s^{\max}=10$. After the search phase is completed, the agent is aligned with the $x$-axis, in a single step, and two additional steps are required to determine the motion vector of the target. Therefore, the total number of discrete steps ($N_{\mathrm{mv}}$) required in the first two phases is given by $N_{\mathrm{mv}} = N_s + 3$. Let $k_L$ denote the first discrete-time instant at which the agent starts updating its position using the computed target motion vector. Since $N_{\mathrm{mv}}=N_s+3$ and $1\leq N_s\leq 10$, the pursuit update begins within $ 4 \leq k_L \leq 13$. Thus, the tracking error becomes constant from $k_L$ step onward.

\section{Simulation}
\noindent The proposed method is evaluated using a custom discrete-time simulator in a two-dimensional plane. The receiver is fixed at $p_R=(0,0)$. The transmitter starts from an arbitrary initial position $p_T(0)$ and moves with a constant velocity $\nu_T$. The position of the mobile agent is initialized to $p_A(0)=(x_A(0),y_A(0))$ and performs search in anti-clockwise direction. The total number of simulation steps is set to $30$. The simulation considers the free-space pathloss model as defined in the system model. The pathloss exponent is set to $\eta=2.0$, which corresponds to free-space propagation, and the reference distance is set to $d_0=1$ meter. Table~\ref{tab:simulation_cases} shows the simulation cases based on the target motion; that is, diagonal motion, axis-parallel motion, different quadrant-coverage patterns, and the failure condition of the proposed method. 

\begin{table}[ht]
\centering
\caption{Simulation cases based on target motion}
\label{tab:simulation_cases}
\renewcommand{\arraystretch}{1.08}
\resizebox{\columnwidth}{!}{%
\begin{tabular}{c c c c}
\hline
Case & $p_T(0)$ & $\nu_T(sign)$ & Purpose \\
\hline
1 & $(10,10)$ & $(0.5,0.5)\;(+,+)$ & One quadrant \\
2 & $(10,5)$ & $(1,-0.5)\;(+,-)$ & Two quadrants \\
3 & $(10,-22)$ & $(-4,4)\;(-,+)$ & Three quadrants(worst case) \\
4 & $(10,10)$ & $(-1,-1)\;(-,-)$ & Through origin \\
5 & $(-20,0)$ & $(4,0)\;(+,0)$ & Along the $x$-axis \\
6 & $(10,5)$ & $(0,1)\;(0,+)$ & Parallel to $y$-axis \\
7 & $(10,1.5)$ & $(0.5,-1)\;(+,-)$ & Failure case \\
\hline
\end{tabular}%
}
\end{table}

The simulation cases are designed using two properties of the target motion: the velocity-sign pattern and the quadrant-coverage pattern. The first four cases cover the four diagonal velocity-sign combinations and the corresponding quadrant-trajectory classes: one-quadrant, two-quadrant, three-quadrant, and origin-passing trajectories. Case 1 represents a one-quadrant trajectory with velocity sign $(+,+)$ as shown in Fig.~\ref{fig:case1}. Here $(+,+)$ denotes that the target moves with increasing x-coordinate and increasing y-coordinate. Here, the target starts in $Q1$ and remains in the same quadrant throughout. This is the baseline condition and fig. ~\ref{fig:case1} verifies the correct performance of the proposed method.
Case 2 represents a two-quadrant trajectory with velocity sign $(+,-)$, as shown in Fig.~\ref{fig:case2}. The target starts in $Q1$ and crosses the $x$-axis, moving from $Q1$ to $Q4$. This case is important because it verifies whether the agent can continue pursuit after the target crosses one coordinate axis and enters an adjacent quadrant. 
Case 3 represents the three-quadrant trajectory with velocity sign $(-,+)$, as shown in Fig.~\ref{fig:case3}. The target starts in $Q4$ and moves through $ Q4 \rightarrow Q3 \rightarrow Q2$. This case represents the worst-case condition as the maximum number of steps required in the search phase is 10. Fig. ~\ref{fig:case3} verifies the worst-case convergence analysis. 
Case 4 represents the trajectory that passes through the origin with velocity sign $(-,-)$, as shown in Fig.~\ref{fig:case4}. Here the target moves diagonally through the origin, transitioning into the diagonally opposite quadrant. This case is important because it tests the method's ability to maintain pursuit even when the target is positioned at $(0,0)$ at some instant. 
Cases 5 and 6 evaluate axis-based target motion, as illustrated in Figs.~\ref{fig:case5} and~\ref{fig:case6}. In Case 5, the transmitter moves only along the $x$-axis with velocity sign $(+,0)$. This case examines target motion along a coordinate axis, where the transmitter lies on one of the coordinate axes instead of one of the quadrants. In Case 6, the transmitter moves parallel to the $y$-axis with velocity sign $(0,+)$. This case evaluates the ability of the proposed method to handle vertical target motion. The previous two cases verify that the proposed method remains valid even when one velocity component is zero, confirming that motion estimation does not depend on having non-zero components along both axes. 
Case 7 represents the failure condition, as shown in Fig.~\ref{fig:case7}. Here, the transmitter changes quadrant after search completes but before computing the motion-vector. This makes the stored quadrant invalid, leading to incorrect estimation of motion vector and failed pursuit. This limitation is proposed to be solved in future.

\begin{figure}[ht]
    \centering
    \includegraphics[width=0.75\linewidth]{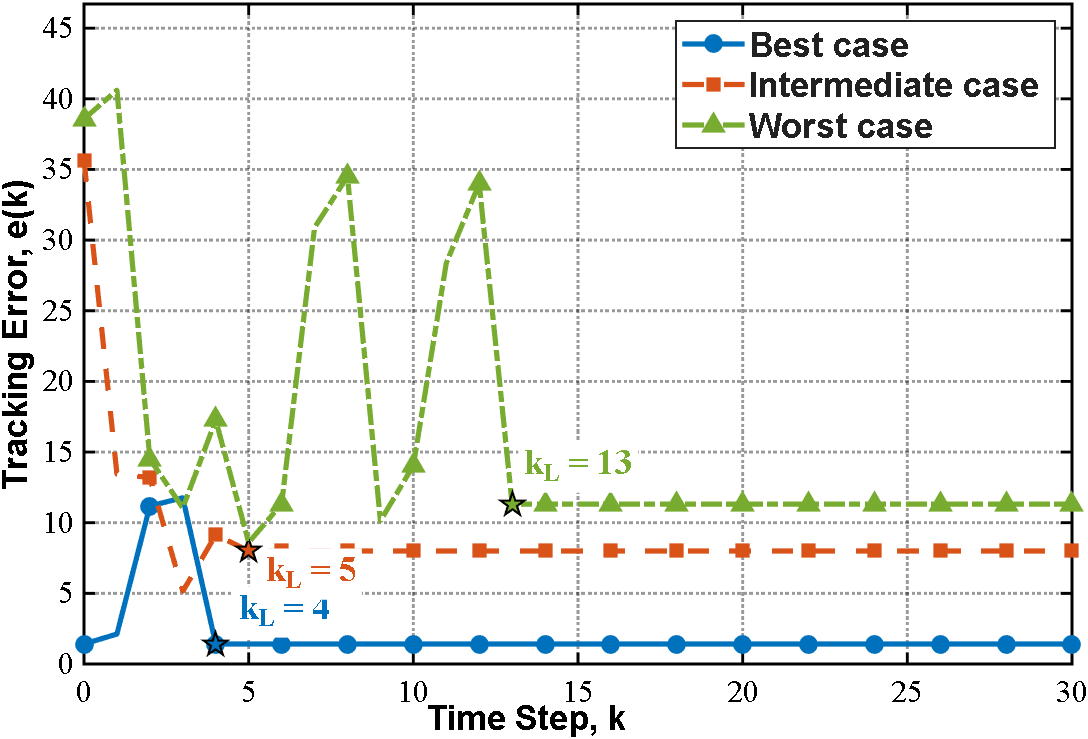}
    \caption{Tracking error $e(k)$ during the search and pursuit phases.}
    \label{fig:error}
\end{figure}

Fig.~\ref{fig:error} shows the tracking error during the search and pursuit stages calculated as shown in (17). The blue curve corresponds to the best case defined in Case 1, where the tracking error becomes constant from $k_L=4$. The orange curve represents the intermediate case defined in Case 5, for which the tracking error becomes constant from $k_L=5$. The green curve denotes the worst case defined in Case 3, where the tracking error becomes constant from $k_L=13$.


\section{Conclusion}
\noindent This paper presents a distance-only search-and-pursuit method for tracking a linearly moving transmitter using one fixed receiver and one mobile agent. The proposed method identifies the target quadrant, estimates its position and motion vector, and guides the agent during pursuit. Under exact ranging and valid quadrant information, the tracking error becomes constant after the motion vector is determined. The proposed method can be extended using an integrated sensing and communication (ISAC) framework, where communication signals are also used to obtain distance, angle-of-arrival, and Doppler information. This can improve target position and motion estimation \cite{guo2025integrated}. Future extensions will consider noisy pathloss measurements, Doppler and angle information from ISAC signals, multi-agent cooperation, and tracking in a three-dimensional coordinate system.


%





\ifCLASSOPTIONcaptionsoff
  \newpage
\fi

\bibliographystyle{IEEEtran}
\bibliography{biblio.bib}

%








\end{document}